\documentclass[english,a4paper,12pt]{article}
\usepackage[T2A]{fontenc}
\usepackage[cp1251]{inputenc}
\usepackage{amsmath}
\usepackage{graphicx}
\usepackage{amssymb}
\usepackage{epsf}
\usepackage{pazh}
\usepackage[labelsep=period,hang]{caption} 

\tightenlines

\newcommand{\nucmu}{\mbox{$m_{\mathrm u}$}}
\newcommand{\WID}[2]{#1_{\mathrm{#2}}}

\newcommand{\DERS}[3]{\left(\frac{\partial #1}{\partial #2}\right)_{\!\!#3}}

\newcommand{\DERPT}[1]{\left(\frac{\partial #1}{\partial
T}\right)_{\!\!\mathrm{pt}}}
\sloppypar

\begin{document}
\baselineskip 21pt


\title{\bf Variational Principle for Stars with a Phase Transition}

\author{\bf \hspace{-1.3cm} \ \
A.V. Yudin\affilmark{1,2*}, T.L. Razinkova\affilmark{1}, D.K. Nadyozhin\affilmark{1,3}}

\affil{ {\it Institute for Theoretical and Experimental Physics,
ul. Bolshaya Cheremushkinskaya 25, Moscow, 117218 Russia}$^1$\\
{\it Novosibirsk State University, ul. Pirogova 2, Novosibirsk, 630090 Russia}$^2$\\
{\it ``Kurchatov Institute'' National Research Center,
pl. Kurchatova 1, Moscow, 123182 Russia}$^3$}

\vspace{2mm}

\sloppypar \vspace{2mm}
\noindent The variational principle for stars with a phase transition has been investigated. The term
outside the integral in the expression for the second variation of the total energy of a star is shown to
be obtained by passage to the limit from the integration over the region of mixed states in the star. The
form of the trial functions ensuring this passage has been found. All of the results have been generalized
to the case where general relativity is applicable. The known criteria for the dynamical stability of a star
when a new phase appears at its center are shown to follow automatically from the variational principle.
Numerical calculations of hydrostatically equilibrium models for hybrid stars with a phase transition have
been performed. The form of the trial functions for the second variation of the total energy of a star that
describes almost exactly the stability boundaries of such stellar models is proposed.


\vfill
\noindent\rule{8cm}{1pt}\\
{$^*$ e-mail: $<$yudin@itep.ru$>$}

\clearpage

\section*{INTRODUCTION}
\noindent The necessity of estimating the stability of a stellar
configuration often arises in various problems of
astrophysics. The variational principle (VP) allows
one to obtain not only the hydrostatic equilibrium
equations for a star from the condition for its total
energy, the sum of its gravitational and internal energies,
being extremal, but also the condition for the
dynamical stability of this equilibrium that ensures
a minimum of the total energy (see Zel’dovich and
Novikov 1967). In this case, the stability condition is
written as the requirement that the second variation of
the integral of the total energy be positive for the entire
set of trial functions describing the various perturbation
modes. However, in the case of a limited number
of trial functions used, the VP gives a necessary but
not sufficient condition for the stability of a star: if the
star is stable for a given perturbation (a given form
of the trial function), then this by no means guarantees
its absolute stability. As experience shows, for
practical purposes it is sometimes sufficient to check
the stability of a star to the simplest perturbations; in
particular, a good approximation for a star without
phase transitions is the investigation of its stability
with respect to homogeneous deformation along the
radius $r$: $\delta r\sim r$. This leads to a well-known stability
condition for the adiabatic index $\gamma$ averaged over the
star: $\langle\gamma\rangle>4/3$, where the averaging is over the mass
coordinate with weight $P/\rho$, while $P$ and $\rho$ are the
pressure and density in the matter, respectively. Thus,
the variational principle is an efficient method for a
practical estimation of the stability of stars.

\section*{THE VARIATIONAL PRINCIPLE IN THE NONRELATIVISTIC REGION}
\noindent Let us first consider the simplest phase transition
(PT): a Maxwellian PT (a typical example is the
liquid–gas transition in homogeneous matter). In
this case, the phase equilibrium conditions lead to the
equality $\DERS{P}{\rho}{T}=0$ in the phase coexistence region.
Let also, for simplicity, the temperature $T=0$. Under
these conditions there is no region of mixed states in
the star, the phases are strictly spatially separated,
and a density jump occurs at the the phase boundary.
The variational principle for a star with such a phase
transition was obtained within the framework of
Newtonian gravity by Bisnovatyi-Kogan et al. (1975)
(see also Bisnovatyi-Kogan 1989):
\begin{equation}
\mathbf{V}=\WID{\mathbf{V}}{I}+\WID{\mathbf{V}}{O}
=\int\limits_0^M\left[P\gamma\rho\left(\frac{d\varphi}{dm}\right)^2-\frac{4\beta}{9}\frac{m\varphi^2}{\upsilon^{7/3}}
\right]dm+\frac{\beta
m_*}{3\upsilon_*^{4/3}}\frac{(\varphi_1-\varphi_2)^2}{\frac{1}{\rho_1}-\frac{1}{\rho_2}}>0,\label{VPmain}
\end{equation}
where $\beta=G(4\pi/3)^{1/3}$ and $\upsilon=4\pi r^3/3$. It is convenient
to use the quantity $\upsilon$ rather than directly $r$
due to the relation $dm=\rho d\upsilon$. The integration in (\ref{VPmain})
is over the mass coordinate $m$ in a hydrostatically
equilibrium star, $M$ is the total mass of the star, $\varphi=\varphi(m)$
is the trial function (different at different phases,
the phases are numbered from the stellar surface)
that describes some perturbation $\delta\upsilon=\varepsilon\varphi(m)$,
$\varepsilon$ is an infinitesimal quantity. As can be seen from (\ref{VPmain}),
in the presence of a phase transition $\mathbf{V}$ is the sum
of two parts: the integral $\WID{\mathbf{V}}{I}$ and outside the integral
$\WID{\mathbf{V}}{O}$. The term outside the integral contains the mass
coordinate $m_*$ and the volume $\upsilon_*$ at the density jump
from $\rho_1$ to $\rho_2$, $\rho_1<\rho_2$. If there are several phase
transitions, then each one has its own corresponding
term outside the integral of the same form.

\subsection*{The Phase Transition at the Stellar Center}
\noindent It follows from the condition for the variations $\delta r,\
\delta\rho$, etc. being bounded that for the central phase
(this is phase 2 in the case of one PT) $\varphi_2(0)=0$. In
this case, the trial functions are not continuous at the
phase boundaries; therefore, if the PT occurs near the
stellar center, then the contribution to $\mathbf{V}$ from phase 1
under the condition $\varphi_1(m_*)\neq 0$ is decisive. Indeed,
the term outside the integral at $m_*\approx\WID{\rho}{2}\upsilon_*$, where
$\rho_2\approx\WID{\rho}{c}$ is the density at the stellar center, tends to
\begin{equation}
\frac{\WID{\rho}{2}}{3\upsilon_*^{1/3}}\frac{\varphi_1^2(0)}{\frac{1}{\rho_1}{-}\frac{1}{\rho_2}}
\end{equation}
and diverges as $1/\upsilon_*^{1/3}$ when $\upsilon_*\rightarrow
0$. The second term, which after the integration is also of
order $O(1/\upsilon_*^{1/3})$, is decisive in the integral. Let us
transform the expression for $\WID{\mathbf{V}}{I}$ so as to gather all
divergences outside the integral. For this purpose, let
us integrate the second term in $\WID{\mathbf{V}}{I}$ twice by parts. We
will then obtain $\mathbf{V}=\WID{\widetilde{\mathbf{V}}}{I}+\WID{\widetilde{\mathbf{V}}}{O}$,
where we separated out the new integral part
\begin{equation}
\WID{\widetilde{\mathbf{V}}}{I}=\int\limits_0^M\left[P\gamma\rho\left(\frac{d\varphi}{dm}\right)^2-
\frac{\beta
m}{3\upsilon^{4/3}}\!\!\left\{\frac{d(\rho\varphi^2)}{dm}+\frac{3\upsilon}{m}\frac{d(\rho^2\varphi^2)}{dm}\right\}\right]dm,
\end{equation}
and the part outside the integral:
\begin{equation}
\WID{\widetilde{\mathbf{V}}}{O}=\frac{\beta}{3\upsilon_*^{4/3}}\!\left\{m_*\frac{(\varphi_1\rho_1{-}\varphi_2\rho_2)^2}{\rho_2-\rho_1}+
3\upsilon_*\left[(\varphi_2\rho_2)^2{-}(\varphi_1\rho_1)^2\right]
\right\}.\label{Vint}
\end{equation}
As before, if there are several PTs, then each one has
its own corresponding contribution to $\mathbf{V}$ of form (\ref{Vint}).
Let now the PT occurs near the center. Then, $\WID{\widetilde{\mathbf{V}}}{I}=O(1)$
and $\WID{\widetilde{\mathbf{V}}}{O}=O(1/\upsilon_*^{1/3})$. Given that $\varphi_2\rightarrow 0$,
$\varphi_1=O(1)$, and $m_*\approx\rho_2\upsilon_*$, we will obtain the stability
condition in a well-known form (see Lighthill 1950):
\begin{equation}
\frac{\rho_2}{\rho_1}<\frac{3}{2}.\label{lambda32}
\end{equation}

\subsection*{The Origin of the Term Outside the Integral}
\noindent The variational principle expressed by Eq. (\ref{VPmain}) is
applicable only for stars in which the phases are
separated spatially and have a well-defined boundary
at which the term outside the integral is calculated.
However, what to do in the case where the
phase coexistence region is present in the star and,
consequently, the density graph has the pattern of a
smoothed step or an even smoother transition? This
can be true both for the Maxwellian description of
the PT, in the case where the star has a nonzero
temperature whose gradient ensures a hydrostatic
equilibrium of the phase coexistence region, and for
the Gibbs description, when $\DERS{P}{\rho}{T}>0$ even
at $T=0$. In this case, the condition for positivity of the
second variation of the star’s energy, from which the
variational principle follows, gives only the integral
part of Eq. (\ref{VPmain}). Let us trace how the term outside
the integral appears when passing to the limit of a
strict spatial phase separation (see the Appendix in
Bisnovatyi-Kogan et al. (1975)).

Thus, let we have a stellar configuration with a
comparatively narrow phase coexistence zone. We
will consider the Maxwellian description of the PT
and assume the zone of mixed states to be described
by an isentrope. This assumption is quite natural:
first, due to the possible action of convection and,
second, due to the presumed narrowness of the spatial
region under consideration. We will then obtain the
limit of cold matter by letting the entropy $S$ approach
zero. Let us separate out the contribution to $\mathbf{V}$ from
the zone of mixed states. The contribution from
the second term in the integrand in (\ref{VPmain}), in view of
its boundedness, approaches zero when $S\rightarrow 0$ and
$\triangle m\rightarrow 0$, where $\triangle m$ is the width of the domain of
integration in the star. Therefore, we can write
\begin{equation}
\triangle\mathbf{V}\approx\int\limits_{\mathrm{mix}}P\gamma\rho\left(\frac{d\varphi}{dm}\right)^2
dm,\label{VPnarrow}
\end{equation}
where the integral is taken over the phase coexistence
region.

Let us now consider the behavior of the parameters
of matter on the isentrope in the region of mixed
states. The phase equilibrium conditions are reduced
to the equality of their pressures and chemical potentials:
\begin{equation}
\left\{
\begin{aligned}
P_1&=P_2, \\
\mu_1&=\mu_2,\label{P_and_mu_PT}
\end{aligned}
\right.
\end{equation}
where the index numbers the phases. The entropy and
density are expressed via the mixing parameter $0\leq\chi\leq 1$
(equal to the mass fraction of the phases) as
\begin{equation}
\begin{aligned}
S&=\chi S_1+(1{-}\chi)S_2, \\
\sigma&=\chi \sigma_1+(1{-}\chi)\sigma_2.\label{S_and_v_PT}
\end{aligned}
\end{equation}
Here, for convenience, we have introduced a quantity
$\sigma\equiv 1/\rho$ that, to within a factor, has the meaning of
volume per unit baryonic charge. For the changes in
these quantities on the isentrope we have
\begin{align}
\delta S&=(S_1-S_2)\delta\chi+\left[\chi\DERPT{S_1}+(1{-}\chi)\DERPT{S_2}\right]\delta T=0 ,\label{DeltaS} \\
\delta\sigma&=(\sigma_1-\sigma_2)\delta\chi+\left[\chi\DERPT{\sigma_1}+(1{-}\chi)\DERPT{\sigma_2}\right]\delta
T.\label{Deltav}
\end{align}
In these expressions the first and second terms describe,
respectively, the change due to the redistribution
of matter between the phases and due to the
change in phase equilibrium conditions. Here, we
have also introduced the notation for the differential
operator
\begin{equation}
\DERPT{}\equiv\DERS{}{T}{P}+\DERPT{P}\DERS{}{P}{T}.
\end{equation}
The quantity $\DERPT{P}$ is found directly from the equilibrium
conditions (\ref{P_and_mu_PT}) (the Clayperon--Clausius formula):
\begin{equation}
\DERPT{P}=\frac{S_1-S_2}{\sigma_1-\sigma_2}.\label{dP_dT_PT}
\end{equation}
Combining Eqs. (\ref{DeltaS}), (\ref{Deltav}) and
(\ref{dP_dT_PT}), we obtain
\begin{equation}
P\rho\gamma=-\DERS{P}{\sigma}{S}=\frac{\DERPT{P}^2}{\chi
\WID{S}{T1}+(1{-}\chi)\WID{S}{T2}}.\label{Prhogamma}
\end{equation}
As can be seen, the quantity $\WID{S}{T}$ introduced here,
\begin{equation}
\WID{S}{T}\equiv\DERPT{S}-\DERPT{\sigma}\DERPT{P}=\DERS{S}{T}{\rho}-
\DERS{\sigma}{P}{T}\left[\DERS{P}{T}{\rho}-\DERPT{P}\right]^2,
\end{equation}
is strictly positive due to the thermodynamic inequalities
\begin{equation}
\DERS{S}{T}{\rho}\geq 0, \quad \DERS{P}{\sigma}{T}\leq 0.
\end{equation}
The hydrostatic equilibrium equation for a star
\begin{equation}
\frac{dP}{dm}=\DERPT{P}\frac{dT}{dm}=-\frac{\beta
m}{3\upsilon^{4/3}},\label{T_and_m_connection}
\end{equation}
gives a relation between $T$ and $m$, while Eq. (\ref{DeltaS}) gives
a relation between $T$ and $\chi$:
\begin{equation}
\DERS{T}{\chi}{S}=-\frac{S_1-S_2}{\chi\DERPT{S_1}+(1{-}\chi)\DERPT{S_2}}.\label{T_and_chi_connection}
\end{equation}
Gathering now Eqs. (\ref{Prhogamma}), (\ref{T_and_m_connection}), and (\ref{T_and_chi_connection}),
given (\ref{dP_dT_PT}), we obtain
\begin{equation}
\triangle\mathbf{V}=\int\limits_0^1\DERS{\varphi}{\chi}{}^{\!\!
2}\left[\frac{\chi\DERPT{S_1}+(1{-}\chi)\DERPT{S_2}}{\chi
\WID{S}{T1}+(1{-}\chi)\WID{S}{T2}}\right]\frac{\beta m \
d\chi}{3\upsilon^{4/3}(\sigma_1{-}\sigma_2)} .\label{VP_delta_f}
\end{equation}
$\WID{S}{T}\rightarrow\DERPT{S}$ when $T\rightarrow 0$, and the expression in
square brackets in (\ref{VP_delta_f}) tends to unity, while all the
remaining quantities, except the term with $\varphi$, can
be taken outside the integral sign, because they are
almost constant in the domain of integration due to
its narrowness. Only the expression
\begin{equation}
\int\limits_0^1\left(\frac{d\varphi}{d\chi}\right)^2
d\chi.\label{dhi2}
\end{equation}
remains under the integral. At fixed values of the trial
function $\varphi_1$ and $\varphi_2$ at the phase boundaries, as is easy
to show, the minimum of the integral (of interest to us)
is ensured by the linear function $\varphi=\varphi_1\chi+\varphi_2(1{-}\chi)$,
while the integral (\ref{dhi2}) itself is equal to $(\varphi_1{-}\varphi_2)^2$, i.e.,
$\triangle \mathbf{V}$ from (\ref{VPnarrow}) turns into the term outside the integral
$\WID{\mathbf{V}}{O}$ from (\ref{VPmain}). Thus, the first term in the integral of
the variational principle when $T\rightarrow 0$ plays the role
of a delta function and, despite the narrowing of the
domain of integration $\triangle m\rightarrow 0$, gives rise to a finite
term outside the integral.

\section*{THE VARIATIONAL PRINCIPLE IN GENERAL RELATIVITY}
\noindent Let us write the stellar equilibrium equations (the
Tolman--Oppenheimer--Volkoff equations) in general
relativity (GR):
\begin{align}
\frac{d P}{d r}&=-\frac{G m(\epsilon+P)}{r^2
c^2}\frac{1+\frac{4\pi P
r^3}{m c^2}}{1-\frac{2 G m}{r c^2}},\label{dPdr_OTO}\\
\frac{d m}{d r}&=4\pi r^2\frac{\epsilon}{c^2}.\label{dmdr_OTO}
\end{align}
Here, $\epsilon$ is the energy of matter per unit volume (including
the rest energy). The condition for the stability
of a star is written in GR via the variational
principle as (see Wheeler et al. 1967; Bisnovatyi-Kogan 1968)
\begin{equation}
\mathbf{V}=4\pi e^{-\Phi(R)}\int\limits_0^R
e^{\Phi(r)}\left[I_1+I_2+I_3\right]dr\geq 0,
\end{equation}
where
\begin{equation}
\Phi(\widetilde{r})=\int\limits_0^{\widetilde{r}}\frac{P+\epsilon}{1-\frac{2
G m}{r c^2}}\frac{4\pi G r}{c^4}dr.
\end{equation}
The terms under the integral are
\begin{equation}
\begin{aligned}
I_1&=\gamma P\left[2\delta r+r\frac{d\delta r}{dr}-\frac{G m}{r
c^2}\frac{1+\frac{4\pi r^3 P}{m c^2}}{1-\frac{2 G m}{r
c^2}}\delta r \right]^2,\\
I_2&=-\frac{P+\epsilon}{\left(1-\frac{2 G m}{r
c^2}\right)^2}\left(1+\frac{4\pi r^3 P}{m
c^2}\right)^2\left(\frac{G m}{r
c^2}\right)^2\delta r^2,\\
I_3&=-\frac{P+\epsilon}{\left(1-\frac{2 G m}{r
c^2}\right)}\left(1+\frac{2\pi r^3 P}{m c^2}\right)\frac{4 G m}{r
c^2}\delta r^2.
\end{aligned}
\end{equation}
As above, let us introduce a variable $\upsilon=4\pi r^3/3$,
$\delta\upsilon=\varepsilon\varphi(m)$ (recall that $\varepsilon$ is an infinitesimal), and
dimensionless combinations
\begin{equation}
p\equiv \frac{P}{\epsilon}, \quad q\equiv \frac{4\pi r^3 P}{m
c^2}, \quad \tau\equiv \frac{G m}{r c^2}.\label{dimensionless}
\end{equation}
Passing to the integration over the mass coordinate,
we then obtain
\begin{equation}
\Phi(\widetilde{m})=\int\limits_0^{\widetilde{m}}\frac{\tau(1+p)}{(1-2\tau)}\frac{dm}{m}.
\end{equation}
The variational integral takes the form
\begin{equation}
\mathbf{V}= e^{-\Phi(M)}\int\limits_0^M
e^{\Phi(m)}\left[\widetilde{I}_1+\widetilde{I}_2+\widetilde{I}_3\right]dm,\label{VP_int_OTO}
\end{equation}
where
\begin{align}
\widetilde{I}_1&=\gamma P\frac{\epsilon}{c^2}\left[\frac{d\varphi}{dm}-\frac{p\tau(1+q)}{q(1-2\tau)}\frac{\varphi}{m} \right]^2,\label{I1}\\
\widetilde{I}_2+\widetilde{I}_3&=-P\frac{\epsilon}{c^2}\left[4+2q+\frac{\tau}{1{-}2\tau}(1+q)^2\right]\frac{\tau
p(1+p)}{1-2\tau}\left(\frac{\varphi}{m q}\right)^2.\label{I23}
\end{align}
It is easy to see that in the nonrelativistic case, $q\ll 1$, $p\ll 1$ and
$\tau\ll 1$, these expressions give the Newtonian
limit (\ref{VPmain}).

\subsection*{The Term Outside the Integral in GR}
\noindent Let us find the form of the term outside the integral
in GR. As in the Newtonian case, it arises from the
integral over the zone of mixed states in the limit
$T\rightarrow 0$. In this case,
\begin{equation}
\triangle\mathbf{V}\approx e^{-\Phi(M)}\int\limits_{\mathrm{mix}}
e^{\Phi(m)} \gamma
P\frac{\epsilon}{c^2}\left(\frac{d\varphi}{dm}\right)^2 dm.
\end{equation}
Repeating the reasoning that led us to Eq. (\ref{VP_delta_f}), with
the only difference that the equilibrium equations are
now given by the relativistic expressions (\ref{dPdr_OTO})--(\ref{dmdr_OTO}),
we will obtain
\begin{equation}
\triangle\mathbf{V}\approx e^{-\Phi(M)}\int\limits_0^1
\frac{e^{\Phi(m)}}{\left(\frac{\partial\sigma}{\partial\chi}\right)_{\!\!
S}}\frac{P+\epsilon}{\rho
c^2}\left(\frac{d\varphi}{d\chi}\right)^2\frac{G m}{4\pi
r^4}\frac{1+\frac{4\pi r^3 P}{m c^2}}{1-\frac{2 G m}{r c^2}}
d\chi.
\end{equation}
Recall that $\sigma\equiv 1/\rho$. For $T\rightarrow 0$
\begin{equation}
\left(\frac{\partial\sigma}{\partial\chi}\right)_{\!\! S}\approx
\frac{1}{\rho_1}-\frac{1}{\rho_2},
\end{equation}
while $\frac{P+\epsilon}{\rho}$ coincides, to within a factor, with the
chemical potential of matter (see Eq. (\ref{mainThermod}) below) and,
hence, is continuous at the phase transition. Therefore,
\begin{equation}
\frac{P+\epsilon}{\rho
\left(\frac{\partial\sigma}{\partial\chi}\right)_{\!\!
S}}=\frac{1}{\frac{1}{P+\epsilon_1}-\frac{1}{P+\epsilon_2}}.
\end{equation}
All of the slowly changing factors can now be taken
outside the integral sign. Repeating the reasoning of
the Newtonian case, we again conclude that the trial
function depends linearly on the mixing parameter in
the phase coexistence region. Finally, for the term
outside the integral $\WID{\mathbf{V}}{O}=\triangle\mathbf{V}$ in GR we have
\begin{equation}
\WID{\mathbf{V}}{O}=\frac{G m_*}{4\pi r_*^4
}\frac{(\varphi_1-\varphi_2)^2}{\left[\frac{c^2}{P_*+\epsilon_1}-\frac{c^2}{P_*+\epsilon_2}\right]}
\frac{1+\frac{4\pi r_*^3 P_*}{m_* c^2}}{1-\frac{2 G m_*}{r_*
c^2}}e^{\Phi(m_*)-\Phi(M)},\label{VneIntOTO}
\end{equation}
where, as above, the symbol $*$ denotes the PT position.
As it must be, we obtain the term outside the
integral from (\ref{VPmain}) in the Newtonian limit. Using the
dimensionless parameters (\ref{dimensionless}) introduced above, this
expression can also be written as
\begin{equation}
\WID{\mathbf{V}}{O}=\frac{P_*^2}{m_*c^2}\frac{(\varphi_1-\varphi_2)^2}{\left[\frac{p_1}{1+p_1}-\frac{p_2}{1+p_2}\right]}
\frac{\tau_*(1+q_*)}{q_*(1-2\tau_*)}e^{\Phi(m_*)-\Phi(M)}.\label{VneIntOTO_bez}
\end{equation}

\subsection*{The Stability Condition at the PT at the Stellar Center}
\noindent Let us derive the dynamical stability condition at
the PT occurring near the stellar center from the variational
principle. According to (\ref{VneIntOTO}), the term outside
the integral diverges as $1/r_*$ when $r_*\rightarrow 0$, because
$\varphi_2\rightarrow 0$ and $\varphi_1\rightarrow \mathrm{const}\neq 0$.
The term $I_3$ makes a
similar contribution in the integral. Near the center
we can write
\begin{equation}
m\approx\frac{4\pi}{3 c^2}\left[\epsilon_2
r_*^3+\epsilon_1\left(r^3-r_*^3\right)\right].
\end{equation}
Retaining only the divergent terms, we will then obtain
\begin{equation}
\WID{\mathbf{V}}{I}\approx-\frac{G\varphi_1^2}{3 c^4
r_*}(\epsilon_1+P_*)(\epsilon_2+3\epsilon_1+6P_*)e^{-\Phi(M)}.
\end{equation}
The term outside the integral in the same approximation is
\begin{equation}
\WID{\mathbf{V}}{O}\approx\frac{G\varphi_1^2}{3 c^4
r_*}\frac{\epsilon_2+3P_*}{\left[\frac{1}{P_*+\epsilon_1}-\frac{1}{P_*+\epsilon_2}\right]}e^{-\Phi(M)}.
\end{equation}
The stability condition $\mathbf{V}=\WID{\mathbf{V}}{I}+\WID{\mathbf{V}}{O}>0$ then immediately
leads to the following relation first derived
by Seidov (1971):
\begin{equation}
\frac{\epsilon_2}{\epsilon_1}<\frac{3}{2}\left(1+\frac{P_*}{\epsilon_1}\right),
\end{equation}
which is a generalization of (\ref{lambda32}) to the case of GR.

\section*{AN EXAMPLE OF APPLYING THE VP}
\begin{figure}[htb]
\epsfxsize=16cm \hspace{-2cm}\center\epsffile{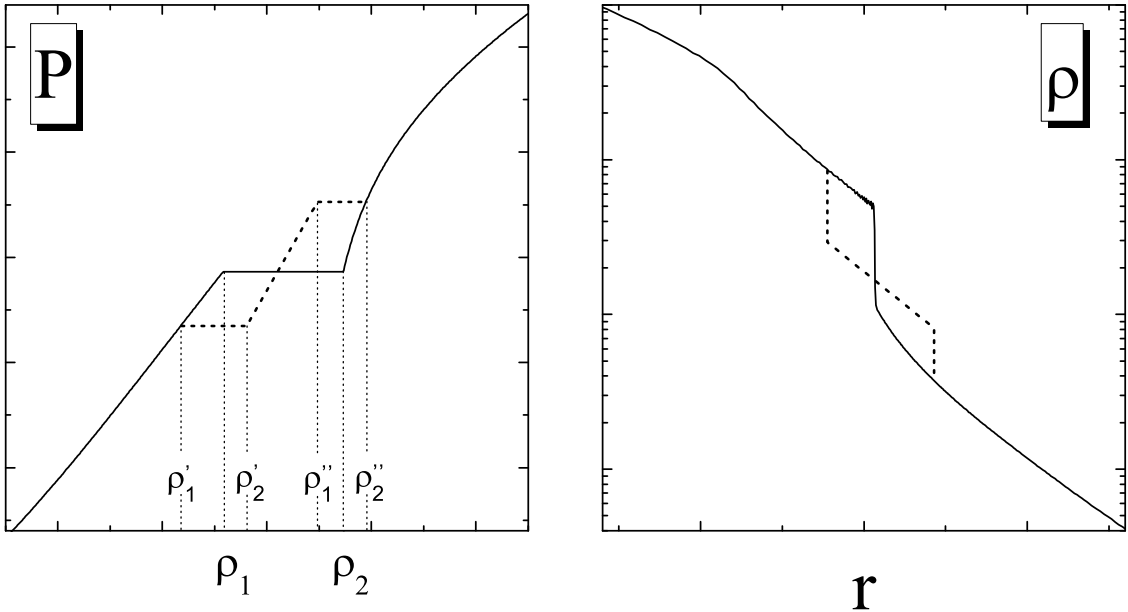}
\caption{\rm Left: dependence $P(\rho)$ for one PT (solid curve) and two PTs (dashed curve line).
Right: the same for the behavior of $\rho(r)$ in
the star.} \label{Pix-Prho}
\end{figure}

\noindent Before turning to the results of our numerical calculations,
let us consider a curious example of applying
the variational principle. For simplicity, we will
work within the framework of Newtonian gravity (the
description in GR is similar). Consider the case of
weak splitting of one PT into two smaller PTs (see
Fig. \ref{Pix-Prho}, the splitting size is exaggerated for clarity).
Let the old values of the PT beginning and end be,
as previously, $\rho_1$ and $\rho_2$, while the new values be $\rho'_1$,
$\rho'_2$ and $\rho''_1$, $\rho''_2$, respectively, with
$\rho'_1\leq\rho_1$, $\rho''_2\geq\rho_2$ and
$\rho'_2\leq\rho''_1$. If the splitting is weak, i.e.,
$\rho'_1\approx\rho_1$ and $\rho''_2\approx\rho_2$,
then the stellar structure changes weakly, while
the regions of the density jumps remain at virtually the
same values $m_*$ and $\upsilon_*$. This means that in Eq. (\ref{VPmain}) for
$\mathbf{V}$ the integral part $\WID{\mathbf{V}}{I}$ remains virtually without any
changes. Omitting the common factor $\frac{\beta m_*}{3\upsilon_*^{4/3}}$, let us
write the term outside the integral $\WID{\mathbf{V'}}{\! O}$ for the case of
two PTs:
\begin{equation}
\WID{\mathbf{V'}}{\! O}\propto
\frac{(\varphi_1{-}\varphi_3)^2}{\frac{1}{\rho_1}-\frac{1}{\rho_2'}}+\frac{(\varphi_3{-}\varphi_2)^2}
{\frac{1}{\rho_1''}-\frac{1}{\rho_2}}\geq
\frac{(\varphi_1{-}\varphi_2)^2}{\frac{1}{\rho_1}-\frac{1}{\rho_2}},
\label{Three_phases}
\end{equation}
where we set $\rho'_1=\rho_1$ and $\rho''_2=\rho_2$ and assume that
the third phase is between the first and second ones.
The last inequality in (\ref{Three_phases}) implies that we always
have $\WID{\mathbf{V'}}{\! O}>\WID{\mathbf{V}}{O}$ for the case under consideration and,
hence, weak splitting of one PT into two smaller PTs
increases the stability margin for the star. This fact
can have important consequences when considering
the stability of hybrid stars, i.e., stars containing ``exotic''
phases inside: quarks etc. It may well be that
the transition to quark matter can occur not immediately
but through a sequence of multi-quark states
(see, e.g., Krivoruchenko et al. (2011) and references
therein). According to what has been said above, this
possibility, if it is realized in nature, can additionally
contribute to the stability of hybrid stars.

\section*{THE CHOICE OF BASIS FUNCTIONS}
\noindent To begin with, we need to choose a form of the trial
function. In doing so, we want to make sure that our
algorithm of using the variational principle is universal
and would be suitable both in the case of a sharp
boundary between the phases in the star (Maxwellian
PT) and in the case of a ``smoothed'' (Gibbs) PT,
where the phases gradually pass into one another and
the region of mixed states is clearly present in the star.
In the most general case, knowing only the equation
of state for matter (i.e., the dependence $P(\rho)$ etc.)
without any information about its phase composition
must be sufficient for us. Thus, we need the functions
common to all phases in the star.

We will seek the trial function $\varphi=\varphi(m)$ as an
expansion in terms of basis function $g_{i}(m)$:
\begin{equation}
\varphi(m)=\sum\limits_{i=1}^{\WID{N}{g}}\alpha_i g_i(m),
\end{equation}
where $\WID{N}{g}$ is the number of basis functions. When this
expression is substituted into the variational principle,
we obtain a stability condition in the form
\begin{equation}
\sum\limits_{i,j=1}^{\WID{N}{g}}M_{i j}\alpha_i \alpha_j\geq 0,
\end{equation}
where the element of the matrix $M_{i j}$ contains both
the contributions from the integrals from $\WID{\mathbf{V}}{I}$ with the
functions $g_i$ and $g_j$ and the contributions from the
term outside the integral $\WID{\mathbf{V}}{O}$ (where it is present).
Consequently, the stability condition is reduced to
the requirement that this quadratic (in coefficients
$\alpha_i$) form be positive definite, which is known to be
equivalent to the condition for positivity of all principal
minors of the matrix $M_{ij}$.

Our main task is now to find the minimal set of
basis functions $g_i$ that would describe best the stability
of stellar configurations with PTs. Undoubtedly,
the list of such functions must include the ``classical''
function $g_1=\upsilon$ that works excellently for stars without
PTs. As follows from the previously considered
method of deriving the term outside the integral in
the VP, the basis function that plays the role of a
delta function in the limit $T\rightarrow 0$ must be linear in
$\chi$ in the region of mixed states. In addition to the
quantities from (\ref{S_and_v_PT}), they also include the internal
energy per unit mass $E$:
\begin{equation}
E=\chi E_1+(1{-}\chi)E_2,
\end{equation}
which is related to the previously used energy per unit
volume $\epsilon$ by the relation $\epsilon=\rho E$
The pressure $P$ and chemical potential $\mu$ experience no jump at the PT,
while the entropy becomes zero for cold configurations.
This means that we have two possibilities: the
basis function must include $E$ or $\sigma=1/\rho$. The relation
between all of the functions listed above is clearly
illustrated by the basic thermodynamic identity
\begin{equation}
E+\frac{P}{\rho}=TS+\frac{Y\mu}{\nucmu},\label{mainThermod}
\end{equation}
where $\nucmu$ is the atomic mass unit, and $Y$ is the
dimensionless concentration. It follows from the condition
for the variations at the stellar center being
bounded that the basis functions must become zero
there, i.e., for example, $E$ must enter into the expression
for the trial function as a combination $E{-}\WID{E}{c}$,
where $\WID{E}{c}$ is its central value. Besides, the basis function
with $\sigma=1/\rho$ must contain the factor removing
the singularity on the stellar surface, for example,
must be $(M{-}m)/\rho$, where $M$ is the total mass of
the star, or $P/\rho-\WID{P}{c}/\WID{\rho}{c}$ etc. In addition, as has
already been noted, the basis function can include
some smoothly changing factor.

\section*{RESULTS OF CALCULATIONS}
\noindent After some numerical experiments, we chose the
following main set of basis functions:
\begin{equation}
\{g_1,g_2,g_3\}=\left\{\upsilon,\frac{M-m}{\rho}-\frac{M}{\WID{\rho}{c}},E-\WID{E}{c}\right\},\label{Nabor_TF}
\end{equation}
with all of the above reservations. However, this
choice is only an example. Here, we set the goal
only to demonstrate the efficiency of the variational
principle. The question about the choice of a minimal
set of optimal trial functions needs to be investigated
further.

\subsection*{The Newtonian Case}
\begin{figure}[h!tb]
\epsfxsize=\textwidth \hspace{-2cm}\center\epsffile{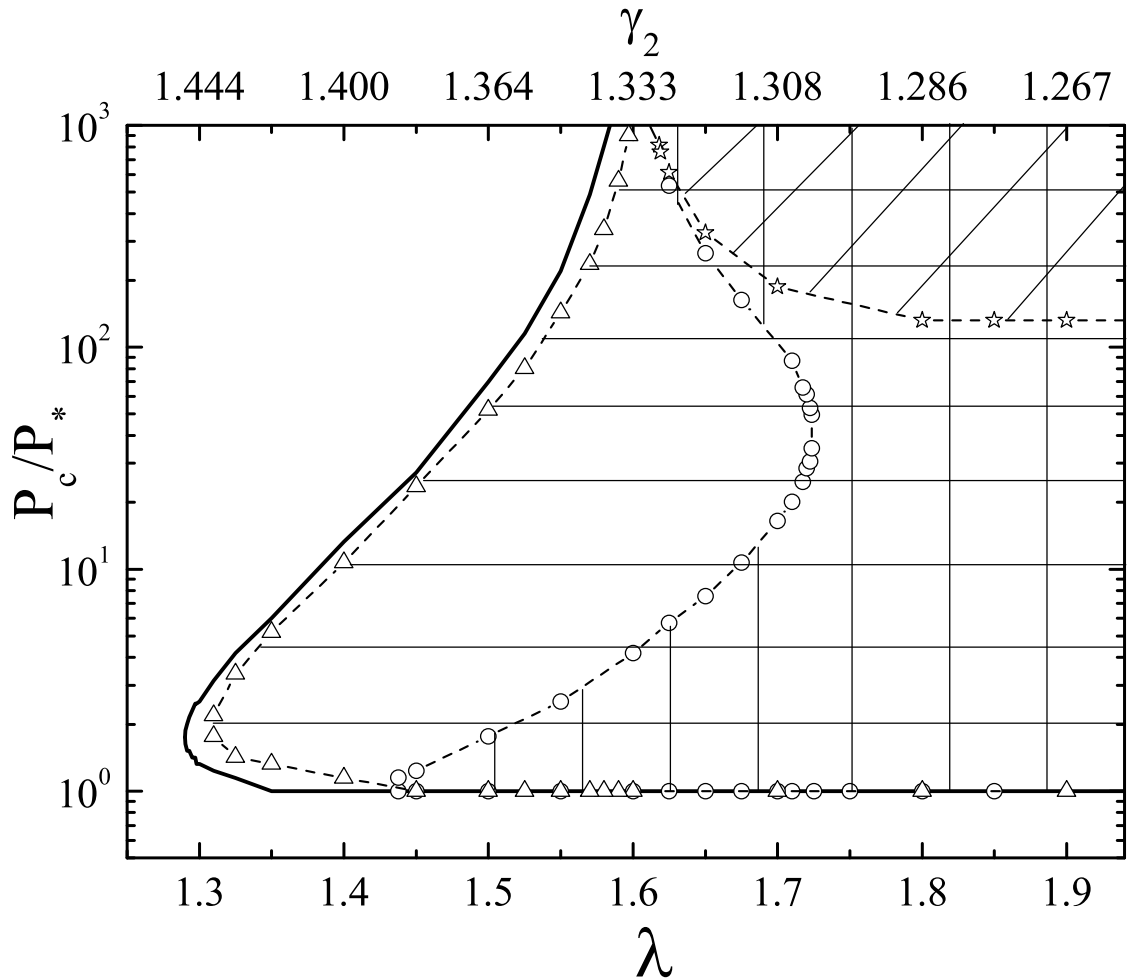}
\caption{\rm The $\left(\frac{\WID{P}{c}}{P_*}, \lambda\right)$
diagram computed within the framework of Newtonian gravity. 
The curves with symbols indicate the
stability boundaries. The oblique, vertical, and horizontal 
hatching indicates the instability zone according to the variational
principle with one, two, and three basis functions, respectively.
 The real instability region is bounded by the thick solid curve.
The values of $\gamma_2$ corresponding to given $\lambda$ are shown at the top.} \label{Pix-Newton}
\end{figure}

\noindent In the nonrelativistic case, the VP is expressed by
Eq. (\ref{VPmain}). To demonstrate how the VP works, we chose
the simplest case of a PT between two polytropes.
Polytrope 1 had an index $n_1=\frac{1}{\gamma_1-1}=3/2$, i.e.,
$\gamma_1=5/3$. The index of the second polytrope $n_2=\frac{1}{\gamma_2-1}$ 
was varied. It is easy to show that the phase
equilibrium conditions (\ref{P_and_mu_PT}) in this case lead to the
following relation between the density jump at the PT
 $\lambda=\rho_2/\rho_1$ and the adiabatic indices:
\begin{equation}
\lambda=\frac{\gamma_2(\gamma_1-1)}{\gamma_1(\gamma_2-1)}.\label{lamb_gam_polytrope}
\end{equation}
Thus, a fixed value of $\lambda$ corresponds to each value
of $\gamma_2$. This makes it possible to compute a one--parameter
sequence of models for hydrostatically
equilibrium stars with various central densities $\rho_c$
and to separate the dynamically stable models from
the unstable ones. For the bipolytropic models with
$\gamma_1=5/3$ and various values of $\gamma_2$ considered here, the
computed boundary between the stable and unstable
models is indicated by the thick solid curve in \ref{Pix-Newton},
which is a diagram: the density jump $\lambda$ is along the
horizontal axis, and the ratio of the central pressure
 $P_c$ to the pressure at the phase transition $P_*$ is along
the vertical axis.
The values of $\gamma_2$ from Eq. (\ref{lamb_gam_polytrope})
corresponding to given $\lambda$ are shown on the upper
axis. The stable and unstable models are located to
the left and the right of the solid curve, respectively.
The stability here was determined by investigating
the behavior of the mass--central density $(M{-}\WID{\rho}{c})$
and mass--radius $(M{-}R)$ curves (see Wheeler
et al. 1967). We will move over the figure from the
bottom upward at fixed $\lambda$. For example, at $\lambda=1.4$
the instability begins immediately when a new phase
appears at $P/P_*=1$ and continues up to $P/P_*\approx 13$,
whereupon the stellar configurations again become
stable. At $\lambda\geq 1.6$ all hybrid configurations are
unstable. Thus, there are several selected density
jumps in the figure: at $\lambda\lesssim 1.29$ all configurations
are stable. The value $\lambda=1.5$ was selected according
to criterion (\ref{lambda32}), while $\lambda>1.6$ correspond, according
to (\ref{lamb_gam_polytrope}), to an adiabatic index of the central phase
$\gamma_2<4/3$.

A digression should be made here: as can be
seen from the figure, our bipolytropic stars lose their
stability with the appearance of a new phase at the
center at $\lambda\approx 1.35$ rather than $\lambda=3/2$, according
to criterion (\ref{lambda32}). However, the contradiction here is
apparent: at $\lambda>3/2$ the loss of stability is guaranteed.
In contrast, at $\lambda<3/2$ the stability will also
depend on the ``stiffness'' of the equation of state for
matter: the ``stiffer'' it is, the greater $\lambda$ (up to the
limit $\lambda=3/2$) is needed to destabilize the star when
a new phase appears at its center. At the same time,
a star with $\gamma=4/3$ at the stability boundary can be
destabilized by an arbitrarily small PT.

Let us now consider the application of the variational
principle. The results of our calculation with
one (first) basis function from set (\ref{Nabor_TF}) are indicated
by the dashed curve with stars. According to the VP,
only the configurations in the upper right corner of the
figure are unstable (the instability region is marked
by the oblique hatching). Such a behavior is quite
understandable: the ``classical'' basis function $\varphi=\upsilon$
does not ``respond'' to a density jump and, in fact,
predicts a stability according to the criterion $\langle\gamma\rangle>4/3$. 
Therefore, an instability in the VP with one basis
function is possible only at $\lambda>1.6$ and a sufficiently
large core of the second phase.

Let us now consider the calculation with two basis
functions (the first and second ones from set (\ref{Nabor_TF})).
The results are indicated by the curve with circles. As
can be seen, the results have improved significantly,
but they are still far from the correct ones. For
example, the instability begins only at $\lambda\approx 1.44$ (the
instability region is marked by the vertical hatching).
Interestingly, the calculation with the first and third
basis functions from set (\ref{Nabor_TF}) gives an even poorer
result. Only the calculation with all three basis functions
simultaneously (indicated by the curve with triangles,
the instability region is marked by the horizontal
hatching) is close to the real state of affairs.

\subsection*{The Calculations in GR}
\noindent In the range of applicability of GR we will consider
two cases as an example. In both cases, we will
use the equation of state from Yudin et al. (2013)
designed to qualitatively model the phase transition
from hadronic matter to quark matter at densities
exceeding the nuclear density $\WID{\rho}{n}\approx
2.6\times 10^{14}~\mbox{g}\cdot\mbox{cm}^{-3}$
by several times. Cold $(T=0)$ matter corresponds to
the first case, the phase transition is Maxwellian, and
the VP includes both the integral part, Eq. (\ref{VP_int_OTO}), and
the term outside the integral (\ref{VneIntOTO_bez}). In the second case,
we consider the stability of isentropes, i.e., stars with
a constant (and comparatively large) entropy per unit
baryonic charge in the matter. Despite the fact that
the PT is still described as a Maxwellian one, as a
result of the presence of a finite temperature gradient
in the matter, the region of mixed states is present in
the star, the boundary between the phases is blurred,
and the VP contains only the integral part (\ref{VP_int_OTO}).

\begin{figure}[htb]
\epsfxsize=\textwidth \hspace{-3cm}\center\epsffile{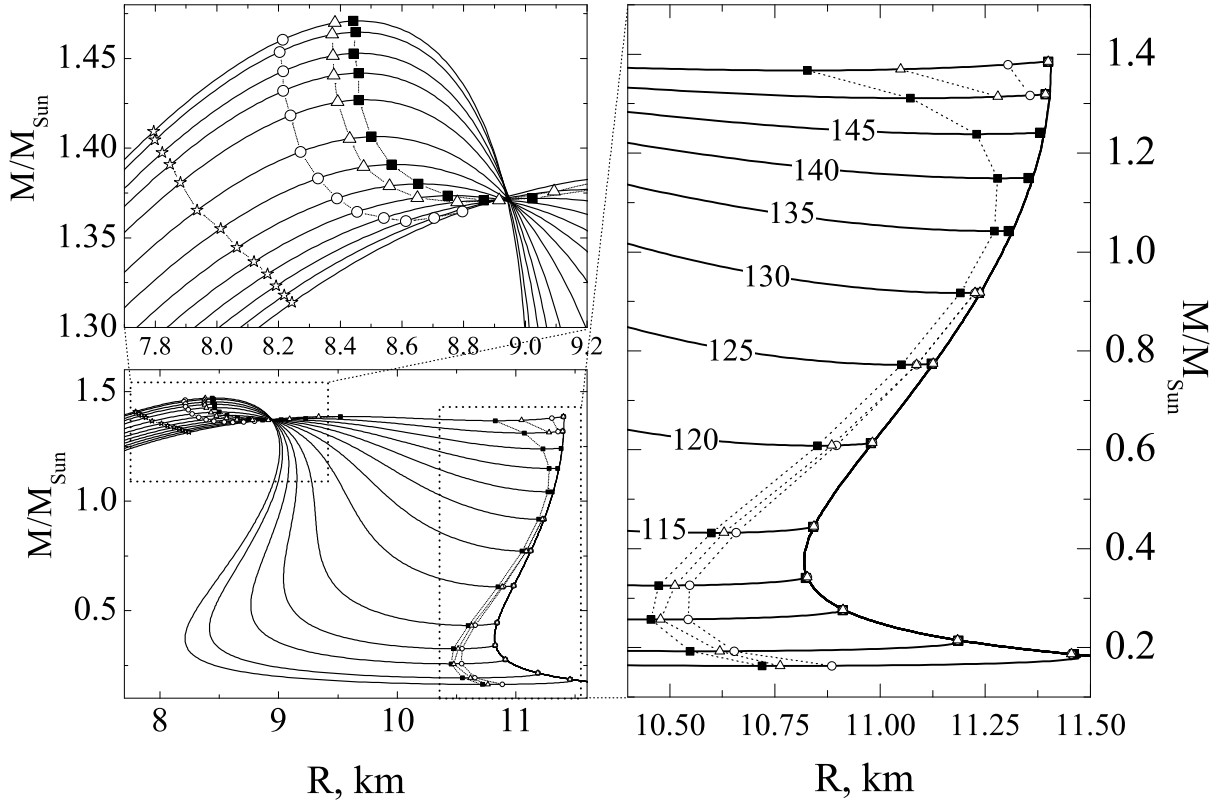}
\caption{\rm Mass--radius $(M{-}R)$ relation for hybrid stars at $\WID{S}{k}=0$. 
Each curve corresponds to its own equation of state for the
quark phase. The values of the bag constant B are indicated by the 
numbers (in units of $\mbox{MeV}\cdot\mbox{fm}^{-3}$) near several curves. The
bottom right panel shows a general view of the $(M{-}R)$ plane; the top left 
and right panels show magnified fragments. For the
remaining explanations see the text.}
\label{Pix-OTO_S0}
\end{figure}
An example of the calculation for the first case
is shown in Fig. \ref{Pix-OTO_S0}. The solid curves in this figure
indicate the mass–radius $(M{-}R)$ relations for hybrid
stars. Each curve corresponds to its own equation of
state for the quark phase (the equations of state for
the hadronic phase are identical). The values of the
bag constant $B$ are indicated by the numbers (in units
of $\mbox{MeV}\cdot\mbox{fm}^{-3}$) near several curves. The bottom left
panel shows a general view of the $(M{-}R)$, plane; the
top left and right panels show magnified fragments.
Moving along the curve from right to left corresponds
to an increase in the central density of the star. At
the instant a new phase appears at the stellar center,
the mass–radius curve abruptly (almost horizontally)
goes to the left of the common envelope representing
the $(M{-}R)$ relation for purely hadronic matter. Different
densities at which quarks appear correspond
to different values of the parameter $B$; the greater
the value of $B$, the higher the density at which quark
matter appears. Such a hybrid star initially becomes
unstable until the core of the new phase becomes
large enough and until the mass--radius curve passes
through the minimum marked by the filled square.
The stable branches of hybrid stars begin from the
points of minimum, which reach the maximum while
passing through the ``singular point'' (i.e., the place
of intersection of the ``bundle'' of $(M{-}R)$ curves corresponding
to different $B$; for an explanation of this
peculiarity, see Yudin et al. 2014). These maxima
correspond to the last stable configurations of hybrid
stars (the black filled squares in the upper left part of
Fig. \ref{Pix-OTO_S0}). As the density at the stellar center increases
further, there are no other stable configurations, and
the star inevitably collapses into a black hole.

Let us now consider how the variational principle
works in this situation. To begin with, we will take
the first, ``classical'' function from our set $g_1=\upsilon$.
For it the boundaries of the regions separating the
stable and unstable models are indicated by the empty
stars. Since this function is continuous at the PT, the
variational principle for it contains only the integral
term. The VP with this function completely ``misses''
the first instability zone shown on an enlarged scale
in the right part of the figure corresponding to the
region between the solid filled squares. However,
the VP with this trial function shows the onset of
instability with a noticeable delay (the stars in the
upper left part of the figure) in the region of high
densities as well. For the model of a star composed
of purely hadronic matter, i.e., without any PT, this
trial function predicts the onset of instability with a
remarkable accuracy: the relative errors in the mass
and radius are $\sim 0.01$ and $0.04\%$, respectively.

Let us now consider the VP with two functions
(open circles) and with the complete set of three
functions (triangles) in Fig. \ref{Pix-OTO_S0}. As can be seen, the
result has improved significantly, the first instability
zone is now resolved, the instability zone beyond the
maxima of the $(M{-}R)$ curves is also considerably
closer to reality, with the results for the complete set
of functions being much better than those for the set
with two functions. The almost horizontal segments
of the mass–radius curves, i.e., the regions in close
proximity to the smooth extrema of the curves, are
the only noticeable discrepancy. This is quite natural:
the almost horizontal segment corresponds to
an indifferent (or nearly indifferent) stellar equilibrium
with respect to perturbations. The border between
stability and instability here is very thin. Thus, we
have shown that the VP with the basis functions of
the specified form is actually capable of predicting
the stability/instability of hybrid stars with a good
accuracy. However, we would like to recall that this
calculation is only an example.

\begin{figure}[htb]
\epsfxsize=16cm \hspace{-2cm}\center\epsffile{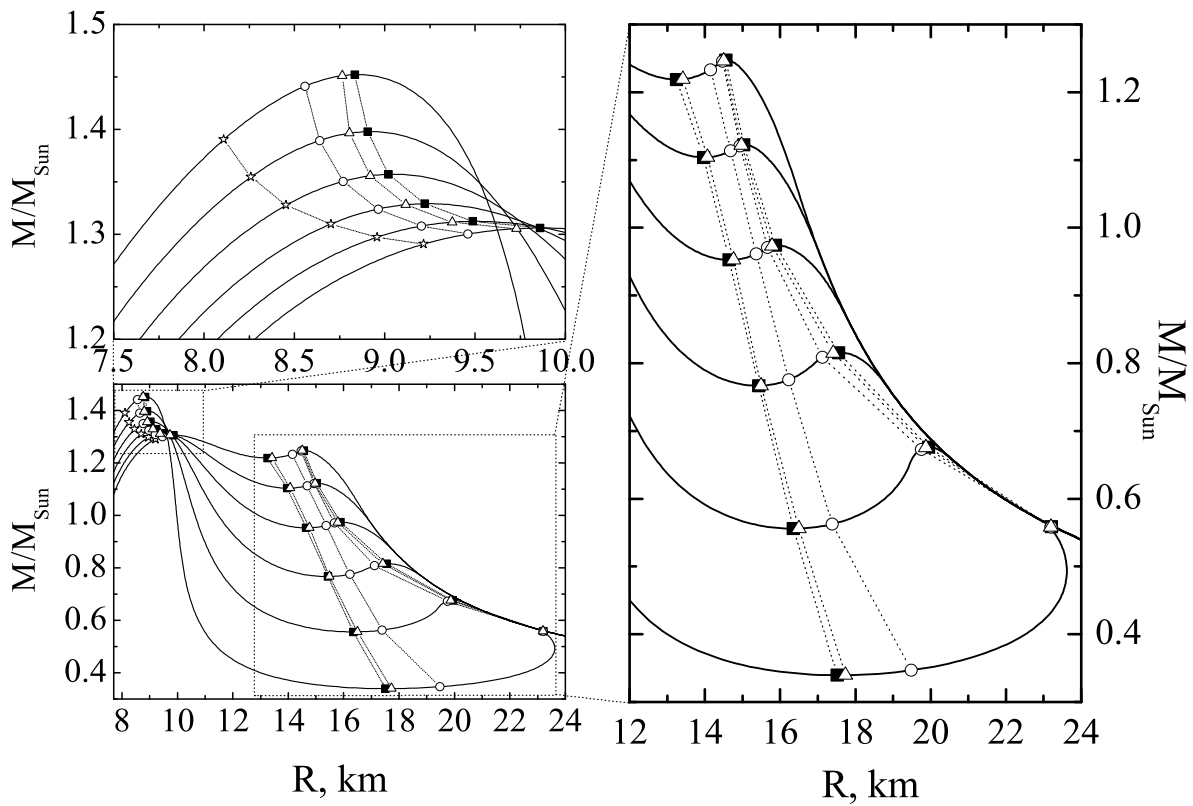}
\caption{\rm Same as Fig.~\ref{Pix-OTO_S0} for
$\WID{S}{k}=2$.} \label{Pix-OTO_S2}
\end{figure}
Let us now turn to the second case. Here, we
consider hot isentropic stellar configurations with dimensionless
entropy $\WID{S}{k}=2$. The results are shown in 
Fig. \ref{Pix-OTO_S2}, where the designations are the same as
those in Fig. \ref{Pix-OTO_S0}. Recall that the VP here works
without the term outside the integral. As can be
seen, the situation is generally similar to the previously
considered case: the VP with one ``classical''
trial function completely misses the first instability
zone at relatively low densities and shows the second
one with a noticeable delay. Note that the situation
with the stability when a new phase appears here is
different from the case of $\WID{S}{k}=0$: now the stability
of a hybrid stellar configuration is lost only when the
core of the new phase will grow to some size rather
than immediately when it appears. On the whole,
however, the results of the work of the VP with two
and especially three basis functions are very close to
the correct description of the stability.

\section*{CONCLUSIONS}
\noindent Let us summarize what has been done in this
paper. We began with the variational principle for
stars with a phase transition that was first obtained
by Bisnovatyi-Kogan et al. (1975). First we demonstrated
that the well-known criterion for the onset
of dynamical instability at PT at the stellar center
$\lambda>3/2$ directly follows from it. We then showed that
the term outside the integral of the VP naturally arises
from the ordinary integral of the variational principle
when using trial functions linear in mixing parameter
in the region of mixed states. These results were
then generalized to the relativistic case, with the form
of the term outside the integral in GR having been
obtained for the first time. Here, we obtained the
generalization of the criterion $\lambda>3/2$ to the case of
GR first found by Seidov (1971) by a different method
directly from the VP.

As a demonstration of the fruitfulness of using
the variational principle, we considered the problem
of weak splitting of one PT into two smaller PTs
and showed that such splitting increases the stability
margin for the star. It would be also interesting to
investigate the case of arbitrarily strong splitting. We
are planning to do this in the immediate future.

Finally, we numerically studied the stability of hybrid
stars within the framework of Newtonian gravity
and in GR. First, we showed that using one ``classical''
basis function $\delta r\propto r$ is quite insufficient to describe
the stability of stars with PT (although without
PT it works excellently in both nonrelativistic and
relativistic cases). Our natural desire would then be
to restrict ourselves to a set of two basis functions
the second of which would belong to the class of
functions linear in mixing parameter in the region of
mixed states that we found. However, it turned out
that only three basis functions, two of which belong to
the above-mentioned class, describe well the stability
of hybrid stars. Since, as has already been said,
each basis function multiplied by a smooth function
that does not become zero at the stellar center can
also serve as a basis one, we cannot be sure that we
actually found the minimal set. We only demonstrated
the fundamental efficiency of the variational principle
in the case of hybrid stars. The problem of searching
for the minimal set of basis functions and their optimal
form requires an additional study.

\clearpage

\section*{ACKNOWLEDGMENTS}
This work was supported by grant
No~11.G34.31.0047 of the Government of the Russian
Federation and SNSF SCOPES grant
no. IZ73Z0-128180/1.


\end{document}